\documentclass[journal]{IEEEtran}
\ifCLASSINFOpdf
\else
\fi
\usepackage{graphicx}
\usepackage{bm}
\usepackage{amsfonts}
\usepackage{amsmath}
\usepackage{amssymb}
\usepackage{times}
\usepackage{subfigure}
\usepackage{latexsym,bm,amsmath,amssymb} 
\usepackage{CJK}
\usepackage{hhline}

\usepackage{multirow} 
\usepackage{amsmath}
\usepackage{xcolor}
\usepackage[linesnumbered,ruled]{algorithm2e}
\usepackage{epstopdf}
\usepackage{algpseudocode}
\usepackage{graphics}
\usepackage{epsfig}
\usepackage{cite}
\usepackage{geometry}
\begin{document}
%
\title{Hybrid MAP and PIC Detection for OTFS Modulation}

\author{Shuangyang~Li,~\IEEEmembership{Student Member,~IEEE,}  Weijie~Yuan,~\IEEEmembership{Member,~IEEE,} Zhiqiang~Wei,~\IEEEmembership{Member,~IEEE,} Jinhong~Yuan,~\IEEEmembership{Fellow,~IEEE,}
Baoming~Bai,~\IEEEmembership{Senior Member,~IEEE}, Derrick Wing Kwan Ng,~\IEEEmembership{Senior Member,~IEEE}, and Yixuan~Xie,~\IEEEmembership{Member,~IEEE}
\vspace{-7mm}
}



\maketitle

\begin{abstract}
Orthogonal time frequency space (OTFS) modulation has attracted substantial attention recently due to its great potential of providing reliable communications in high-mobility scenarios. In this paper, we propose a novel hybrid signal detection algorithm for OTFS modulation.
By characterizing the input-output relationship of OTFS modulation, we derive the near-optimal \emph{symbol-wise} maximum \emph{a posteriori} (MAP) detection algorithm for OTFS modulation,
which aims to extract the information of each transmitted symbol based on the corresponding related received symbols.
Furthermore, in order to reduce the detection complexity, we propose a partitioning rule that separates the related received symbols into two subsets for detecting each transmitted symbol, according to the corresponding path gains. We then introduce a hybrid detection algorithm to exploit the power discrepancy of each subset, where the MAP detection is applied to the subset with larger channel gains, while the parallel interference cancellation (PIC) detection is applied to the subset with smaller channel gains. Simulation results show that
the proposed algorithms can not only approach the performance of the near-optimal \emph{symbol-wise} MAP algorithms, but also offer a substantial performance gain compared with existing algorithms.
\end{abstract}

\begin{IEEEkeywords}
Orthogonal time frequency space (OTFS), reduced-complexity detection, sum-product algorithm.
\vspace{-2mm}
\end{IEEEkeywords}

%
\IEEEpeerreviewmaketitle

\section{Introduction}
Various emerging applications, such as mobile communications on board aircraft (MCA), low-earth-orbit satellites (LEOSs), high speed trains, and unmanned aerial vehicles (UAVs) \cite{Meyer2019road,cai2020joint}, are expected to operate in high-mobility environments, which imposes great challenges for next generation wireless communications.
However, the currently deployed orthogonal frequency division multiplexing (OFDM) modulation is very vulnerable to the severe inter-carrier interference
due to the significant Doppler spread introduced by the high-mobility~\cite{hwang2008ofdm}.

The recently proposed orthogonal time frequency space (OTFS) modulation
provides a potential solution for reliable communications in high-mobility scenarios \cite{Hadani2017orthogonal}.
Different from the conventional OFDM modulation, OTFS modulation places the information symbols in the delay-Doppler (DD) domain instead of the time-frequency (TF) domain.
It can be shown that with the DD domain data multiplexing, each transmitted symbol principally experiences the whole fluctuations of the TF channel over an OTFS frame. Thus, OTFS modulation offers the potential of exploiting the full channel diversity, achieving a better error performance compared with that of the conventional OFDM modulation in high-mobility environments~\cite{Hadani2017orthogonal}.


In order to achieve the potential full channel diversity, advanced detection methods are required for OTFS detection.
The \emph{symbol-wise} maximum \emph{a posteriori} (MAP) detection is the optimal detection method in the sense of minimizing the bit error rate (BER) but it usually requires an exceedingly high detection complexity which increases exponentially with the number of paths of the channel. As a compromised approach, a messaging passing algorithm was proposed in \cite{Raviteja2018interference}, where Gaussian approximation is applied to model the characteristic of interferences. Nevertheless, simply treating all interferences as Gaussian variables may introduce substantial performance loss relative to the MAP detection.


In this paper, we propose a novel hybrid MAP and
parallel interference cancellation (PIC) detection method for OTFS modulation with a reduced computational complexity.
Based on the framework of sum-product algorithm \cite{Kschischang2001factor}, we first derive the near-optimal detection algorithm in the \emph{symbol-wise} MAP sense,
which needs to consider all possible combinations of the related received symbols in the DD domain, for detecting each transmitted symbol.
Note that the detection complexity of this method is exponential to the number of independent paths, which becomes prohibitively high for a large number of paths.
To reduce the detection complexity of MAP detection, we further propose a partitioning rule for the related received symbols based on the path gains of the channel, where
the related symbols are separated into two subsets. Thus, a hybrid detection algorithm is naturally introduced to exploit the power discrepancy of each subset.
On one hand, the symbols from the subset corresponding to small path gains are approximated as Gaussian random variables based on the \emph{a prior} mean and variance and are cancelled by performing PIC; On the other hand, the interference induced by the symbols from the subset with large path gains is regarded as \emph{useful} information that is effectively extracted by the MAP detection. The detection complexity of the proposed hybrid detection algorithm is only exponential to the size of subset with large path gains.
More importantly, the proposed hybrid detection algorithm offers the flexibility to adapt different detection parameters based on the channel condition, thereby providing a good trade-off between the detection performance and the complexity.
Simulation results show that the proposed hybrid MAP and PIC detection outperforms the existing OTFS detection algorithm and only has a marginal performance loss (less than 1 dB) to the near-optimal \emph{symbol-wise} MAP algorithm.

\emph{Notations:} We use ${\mathbb{A}}$ to denote the signal constellation and ${\mathbb{E}}$ to denote the expectation operation, respectively;
We use ${\left[ {\cdot} \right]_N}$ to denote the modulo $N$ operation; ${{{\bf{F}}_N}}$ and ${{{\bf{I}}_M}}$ denote the discrete Fourier transform (DFT) matrix of size $N\times N$ and the identity matrix of size $M\times M$, respectively; $\delta(\cdot)$ denotes the Dirac delta function; $\textrm{vec}(\cdot)$ denote the vectorization operation;
$(\cdot)^{*}$ denote the conjugate operation; $\propto$ represents both sides of the equation are multiplicatively connected to a constant;
$\Pr(\cdot)$ denotes the probability of an event.

\section{System Model}
\begin{figure}
\centering
\includegraphics[width=3.5in,height=1.5in]{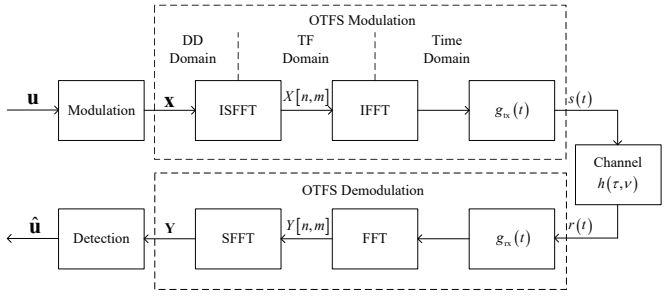}
\caption{The block diagram of the considered OTFS system.}
\vspace{-2mm}
\label{System_Model}
\centering
\end{figure}
Without loss of generality, let us consider an OTFS system whose block diagram is given in Fig. \ref{System_Model}.
Let $N$ be the number of time slots and $M$ be the number of sub-carriers for each OTFS symbol, respectively.
An information sequence $\bf{u}$ is modulated into ${\bf{x}} \in {{\mathbb{A}}^{M N}}$ with length $MN$.
In particular, the information symbol vector ${\bf{x}}$ can be arranged as a two-dimensional (2D) matrix ${\bf{X}} \in {{\mathbb{A}}^{M \times N}}$, i.e., ${\bf{x}} \buildrel \Delta \over = \textrm{vec}\left( {\bf{X}} \right)$, and the $(k,l)$-th element of ${\bf{X}}$, $x\left[ {k,l} \right]$, is the modulated signal in the $k$-th Doppler and $l$-th delay grid \cite{Hadani2017orthogonal}, for $0 \le k \le N-1,0 \le l \le M-1$.
The TF domain transmitted symbol $X\left[ {n,m} \right], 0 \le n \le N-1, 0 \le m \le M-1$ is obtained according to ${\bf{X}}$ via the inverse symplectic finite Fourier transform (ISFFT) \cite{Hadani2017orthogonal}, i.e.,
\begin{equation}
X\left[ {n,m} \right] = \frac{1}{{\sqrt {NM} }}\sum\limits_{k = 0}^{N - 1} {\sum\limits_{l = 0}^{M - 1} {x\left[ {k,l} \right]} } {e^{j2\pi \left( {\frac{{nk}}{N} - \frac{{ml}}{M}} \right)}}  .
\end{equation}
The time domain OTFS signal $s\left( t \right)$ can be obtained by sending ${\bf{X}}$ to a conventional OFDM modulator, which is written as
\begin{equation}
s\left( t \right) = \sum\limits_{n = 0}^{N - 1} {\sum\limits_{m = 0}^{M - 1} {X\left[ {n,m} \right]{g_{{\rm{tx}}}}\left( {t - nT} \right){e^{j2\pi m\Delta f\left( {t - nT} \right)}}} },
\end{equation}
where ${\Delta f}$ is the frequency spacing between adjacent sub-carriers, $T$ is the time slot duration, i.e., $T\Delta f = 1$, and $g_{{\rm{tx}}}(t)$ is the transmitter shaping pulse.

Similar to \cite{Hadani2017orthogonal}, we consider the DD domain representation of the time-varying channel, where the channel impulse response is given by
\begin{equation}
h\left( {\tau ,\nu } \right) = \sum\limits_{i = 1}^P {{h_i}\delta \left( {\tau  - {\tau _i}} \right)\delta \left( {\nu  - {\nu _i}} \right)}.
\label{channel}
\end{equation}
In (\ref{channel}), $P$ is the number of paths and $h_i$, $\tau _i$, and $\nu _i$ are the path gain, delay, and Doppler shift corresponding to the $i$-th path, respectively.
Specifically, we denote by $l_\tau ^{\left( i \right)}$ and $l_\nu ^{\left( i \right)}$ the indices of delay and Doppler, respectively, where we have
\begin{equation}
{\tau _i} = \frac{{l_\tau ^{\left( i \right)}}}{{M\Delta f}},
{\nu _i} = \frac{{l_\nu ^{\left( i \right)}}}{{NT}}.
\label{resolution}
\end{equation}
Note that in (\ref{resolution}), the terms $\frac{1}{{M\Delta f}}$ and $\frac{1}{{NT}}$ refer to the delay and Doppler \emph{resolutions}, respectively \cite{Raviteja2018interference}.
For simplicity, in this paper, we only consider the case where both ${l_\tau ^{\left( i \right)}}$ and ${l_\nu ^{\left( i \right)}}$ are integers, i.e., the OTFS system does not have fractional delay or Doppler shifts \cite{Raviteja2019effective,Li2020performance}.
We note that the fractional delay and Doppler shifts can be addressed by adding virtual integer taps in the DD domain channel \cite{fish2013delay} or by applying TF domain windows \cite{Wei2020transmitter}{\footnote{In practice, the
non-fractional case can be achieved by using sufficiently large $M$ and $N$ \cite{Raviteja2018interference}.}}.
Furthermore, we assume that the path gain follows the Rayleigh distribution with respect to the exponential power delay profile \cite{hlawatsch2011wireless}, where the variance $\hat {\sigma}_i^2$ corresponding to the $i$-th path gain satisfies
\begin{equation}
\hat {\sigma}_i^2 = \frac{{\exp \left( { - {{{l_\tau ^{\left( i \right)}}} \mathord{\left/
 {\vphantom {{{l_\tau ^{\left( i \right)}}} {10}}} \right.
 \kern-\nulldelimiterspace} {10}}} \right)}}{{\sum\limits_{k = 1}^P {\exp \left( { - {{{l_\tau ^{\left( i \right)}}} \mathord{\left/
 {\vphantom {{{l_\tau ^{\left( i \right)}}} {10}}} \right.
 \kern-\nulldelimiterspace} {10}}} \right)} }}.
\end{equation}


At the receiver side, following the conventional OFDM demodulation, the TF domain received symbols $Y\left[ {n,m} \right]$ can be written as
\begin{equation}
Y\left[ {n,m} \right] = \sum\limits_{n' = 0}^{N - 1} {\sum\limits_{m' = 0}^{M - 1} {{H_{n,m}}\left[ {n',m'} \right]} } X\left[ {n',m'} \right]+w \left[ {n,m} \right],
\label{receivded_symbols_TF}
\end{equation}
where $w \left[ {n,m} \right]$ is the corresponding TF domain white noise sample with one-sided power spectral density (PSD) is $N_0$, and
\begin{align}
&{H_{n,m}}\left[ {n',m'} \right] \notag\\
=& \int {\int {h\left( {\tau ,\nu } \right){A_{{g_{{\rm{tx}}}},{g_{{\rm{rx}}}}}}\left( {\left( {n - n'} \right)T - \tau ,\left( {m - m'} \right)\Delta f - \nu } \right)} } \notag \\
&{e^{j2\pi \left( {\nu  + m'\Delta f} \right)\left( {\left( {n - n'} \right)T - \tau } \right)}}{e^{j2\pi \nu n'T}}d\tau d\nu,
\label{TF_channel_coef}
\end{align}
denotes the corresponding channel in the TF domain.
In (\ref{TF_channel_coef}), the function ${A_{{g_{{\rm{tx}}}},{g_{{\rm{rx}}}}}}\left( {{\tau _\Delta },{\nu _\Delta }} \right)$ is the so-called cross-ambiguity function, which is given by \cite{Raviteja2018interference}
\begin{equation}
{A_{{g_{{\rm{tx}}}},{g_{{\rm{rx}}}}}}\left( {{\tau _\Delta },{\nu _\Delta }} \right) \buildrel \Delta \over = \int {{g_{{\rm{tx}}}}\left( t \right)g_{{\rm{rx}}}^*} \left( {t - {\tau _\Delta }} \right){e^{j2\pi {\nu _\Delta }\Delta ft}}dt,
\end{equation}
where ${g_{{\rm{rx}}}}\left( t \right)$ is the receiver filter.
For simplicity, we only consider the case that ${A_{{g_{{\rm{tx}}}},{g_{{\rm{rx}}}}}}\left( {{\tau _\Delta },{\nu _\Delta }} \right) = \delta \left( {{\tau _\Delta }} \right)\delta \left( {{\nu _\Delta }} \right)$, i.e., the pulses of the transmitter and receiver are \emph{ideal} such that the bi-orthogonal condition \cite{Raviteja2018interference} holds.
Note that although ideal pulses are not practically realizable, they can be well approximated by waveforms with a support concentrated
as much as possible in time and in frequency \cite{Raviteja2018interference}.
Moreover, the proposed detection algorithm can be straightforwardly extended to the cases of non-ideal pulses, such as the rectangular pulse case.
Based on the bi-orthogonal assumption, (\ref{receivded_symbols_TF}) can be simplified as
\begin{equation}
Y\left[ {n,m} \right] = H\left[ {n,m} \right]X\left[ {n,m} \right] + w \left[ {n,m} \right],
\end{equation}
where
\begin{equation}
H\left[ {n,m} \right] = \int {\int {h\left( {\tau ,\nu } \right)} } {e^{ - j2\pi \left( {\nu  + m'\Delta f} \right)\tau }}{e^{j2\pi \nu nT}}d\tau d\nu .
\end{equation}
By applying the SFFT, the overall input-output relationship between the DD domain transmitted symbols $x\left[ {k,l} \right]$ and received symbols $y\left[ {k,l} \right]$ is given by \cite{Raviteja2018interference}
\begin{equation}
y\left[ {k,l} \right] = \sum\limits_{i = 1}^P {{h_i}{e^{ - j2\pi {\nu _i}{\tau _i}}}x\left[ {{{\left[ {k - l_\nu ^{\left( i \right)}} \right]}_N},{{\left[ {l - l_\tau ^{\left( i \right)}} \right]}_M}} \right]}  + \eta \left[ {k,l} \right],
\label{DD_model}
\end{equation}
where $\eta \left[ {k,l} \right]$ is the corresponding white noise sample in the DD domain.
Similar to $\bf{X}$, we denote by $\bf{Y}$ the 2D received symbols, whose $(k,l)$-th element is
$y\left[ {k,l} \right]$.

Without loss of generality, we design the data detection algorithm based on (\ref{DD_model}) in the sequel.

\section{Symbol-Wise MAP Detection for OTFS Modulation}
In this section, we derive the symbol-wise MAP Detection for OTFS Modulation. Although this derivation is straightforward, it has not been introduced in the literature of OTFS modulation to the best of the knowledge of authors.
The detection can be carried out based on the \emph{symbol-wise} maximum \emph{a posterior} (MAP) rule, i.e.,
\begin{equation}
{\hat x}\left[ {k,l} \right] = \arg \mathop {\max }\limits_{x\left[ {k,l} \right] \in {\mathbb A}} \Pr \left\{ {x\left[ {k,l} \right]|{\bf{Y}}} \right\},
\label{MAP_rule}
\end{equation}
where ${\hat x}\left[ {k,l} \right]$ is the element at the $k$-th row and $l$-th column in the 2D estimated symbol matrix ${\bf{\hat X}}$.
For notational brevity, let us define the following sets.
\begin{align}
\mathbb{H}^{\left( i \right)}&\buildrel \Delta \over = \left\{ {{h_j}\left| {1 \le j \le P,j \ne i} \right.} \right\},\notag\\
\mathbb{Y}_{k,l} &\buildrel \Delta \over =  \left\{ {y\left[{{\left[ {k + l_\nu ^{\left( i \right)}} \right]}_N},{{\left[ {l + l_\tau ^{\left( i \right)}} \right]}_M}\right]\big| {1 \le i \le P} } \right\},\notag\\
\mathbb{X}_{k,l}^{\left( i \right)}&\!\buildrel \Delta \over = \!\!
 \left\{\! {x\!\left[ {{{\left[ {k \!+\! l_\nu ^{\left( i \right)} \!-\! l_\nu ^{\left( j \right)}} \right]}\!_N}\!,\!{{\left[ {l \!+\! l_\tau ^{\left( i \right)} \!-\! l_\tau ^{\left( j \right)}} \right]}\!_M}} \right]\!\!\big| {1 \le j \le P}\!\! ,j \ne i} \right\}.\notag
\end{align}
According to~\eqref{DD_model}, it can be shown that the set $\mathbb{Y}_{k,l}$ contains the $P$ received symbols that are associated to the DD domain transmitted symbol $x\left[ {k,l} \right]$,
while the set $\mathbb{X}_{k,l}^{\left( i \right)}$ contains $P-1$ DD domain transmitted symbols that are related to the received symbol ${\mathbb{Y}_{k,l}}\left[ i \right]$.
In particular, the probability $\Pr \left\{ {x\left[ {k,l} \right]|{\bf{Y}}} \right\}$ can be factorized with respect to $\mathbb{Y}_{k,l}$ and $\mathbb{X}_{k,l}^{\left( i \right)}$, for which we have the following Theorem.

\textbf{Theorem 1} \emph{(Probability factorization)}:
Assuming that the transmitted symbols in $\bf{X}$ are independently taking values in the constellation set ${\mathbb{A }}$ with equal probabilities, the \emph{a posteriori} probability of $\Pr \left\{ {x\left[ {k,l} \right]|{\bf{Y}}} \right\}$ can be approximated as
\begin{align}
\Pr \left\{ {x\left[ {k,l} \right]|{\bf{Y}}} \right\} \approx & \prod\limits_{i = 1}^P {\sum\limits_{{\mathbb{X}_{k,l}^{\left( i \right)}}} {\Pr } } \left\{ { {\mathbb{Y}_{k,l}}\left[ i \right]\left| {{\mathbb{X}_{k,l}^{\left( i \right)}},x\left[ {k,l} \right]} \right.} \right\} \notag\\
&\Pr \left\{ {{\mathbb{X}_{k,l}^{\left( i \right)}}\left| {{{\bf{Y}}_{ \notin {\mathbb{Y}_{k,l}}\left[ i \right]}}} \right.} \right\}\Pr \left\{ {x\left[ {k,l} \right]} \right\},
\label{sum_product}
\end{align}
where ${{\bf{Y}}_{ \notin {\mathbb{Y}_{k,l}}\left[ i \right]}}$ denotes the set of $\bf{Y}$ excluding the element ${\mathbb{Y}_{k,l}}\left[ i \right]$.

\emph{Proof}: The proof follows the standard sum-product algorithm and is given in the Appendix.

\begin{figure}
\includegraphics[width=3.5in,height=1.3in]{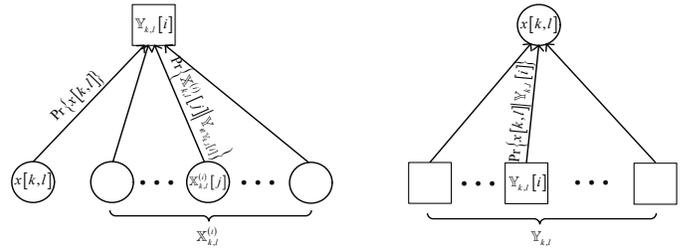}
\caption{The probabilistic graphical model of the \emph{symbol-wise} MAP algorithm.}
\label{Factor_graph1}
\centering
\vspace{-3mm}
\end{figure}

It can be observed that the probability factorization given in (\ref{sum_product}) can be fully characterized by a probabilistic graphical model
and the approximation in~\eqref{sum_product} becomes exact when the corresponding model does not contain any cycles.
Furthermore, the probability factorization in~\eqref{sum_product}
can be efficiently implemented by using the sum-product algorithm \cite{Kschischang2001factor}, where the messages are passed according the corresponding graphical model and are updated with respect to the update rule of function nodes.
Without loss of generality, the considered graphical model is given in Fig. \ref{Factor_graph1}. As the figure implies, the \emph{a prior} probability $\Pr\left\{{x\left[ {k,l} \right]}\right\}$ is passed from the variable node $x\left[ {k,l} \right]$ to the function node ${\mathbb{Y}_{k,l}}\left[ i \right]$; For each variable node ${\mathbb{X}_{k,l}^{\left( i \right)}}[j]$, the probability $\Pr \left\{ {{\mathbb{X}_{k,l}^{\left( i \right)}[j]}\left| {{{\bf{Y}}_{ \notin {\mathbb{Y}_{k,l}}\left[ i \right]}}} \right.} \right\}$ is passed to the function node ${\mathbb{Y}_{k,l}}\left[ i \right]$; On the other hand, for each function node ${\mathbb{Y}_{k,l}}\left[ i \right]$, the probability $\Pr \left\{ {\left. {x\left[ {k,l} \right]} \right|{\mathbb{Y}_{k,l}}\left[ i \right]} \right\}$ is passed to the variable node $x\left[ {k,l} \right]$. Specifically, we have
\begin{align}
&\Pr \left\{ {\big. {{\mathbb{Y}_{k,l}}\left[ i \right]} \big|\mathbb{X}_{k,l}^{\left( i \right)},x\left[ {k,l} \right]} \right\} = \frac{1}{{\sqrt {\pi {N_0}} }}\notag\\
&\exp \left( {\! -\! {{\left| {{\mathbb{Y}_{k,l}}\left[ i \right] \!-\! \sum\limits_{\scriptstyle j = 1\hfill\atop
}^{P-1} {{\mathbb{H}^{\left( i \right)}[j]}\mathbb{X}_{k,l}^{\left( i \right)}\left[ j \right] \!- \!{h_i}x\left[ {k,l} \right]} } \right|}^2}\!\Bigg /\!\!{N_0}} \right),
\label{probability1}
\end{align}
and
\begin{equation}
\Pr \left\{ {x\left[ {k,l} \right]\left| {{{\bf{Y}}_{ \notin {\mathbb{Y}_{k,l}}\left[ i \right]}}} \right.} \right\} \propto \prod\limits_{\scriptstyle j = 1\hfill\atop
\scriptstyle j \ne i\hfill}^P {\Pr \left\{ {\left. {x\left[ {k,l} \right]} \right|{\mathbb{Y}_{k,l}}\left[ j \right]} \right\}} .
\label{probability2}
\end{equation}
The detailed procedures for the \emph{symbol-wise} MAP algorithm are summarized in Algorithm~1.

\begin{algorithm}
\caption{Symbol-Wise MAP Detection Algorithm for OTFS Modulation}
\KwIn{$\bf{Y}$, ${\mathbb{A}}$, $M$, $N$, $P$, the maximum number of iteration $I_{\max}$, \emph{a prior} probability $\Pr\left\{{x\left[ {k,l} \right]}\right\}$ and the channel state information $h_i$, ${l_\nu ^{\left( i \right)}}$, ${l_\tau ^{\left( i \right)}}$, for $1\le i \le P$.}
\KwOut{${\bf{\hat X}}$ and $\Pr \left\{ {\left. {x\left[ {k,l} \right]} \right|\bf{Y}} \right\}$.}
\For {$I=1;I \le I_{\max }$}
{
    \For {$i=1;i \le P$}
    {
        \For {$k=0;k \le N-1$}
        {
            \For {$l=0;l \le M-1$}
            {
                Enumerate all combinations of ${\mathbb{X}_{k,l}^{\left( i \right)}}$.\\
                For each possible combination of ${\mathbb{X}_{k,l}^{\left( i \right)}}$, compute
                (\ref{probability1}) and $\Pr \left\{ {{\mathbb{X}_{k,l}^{\left( i \right)}[j]}\left| {{{\bf{Y}}_{ \notin {\mathbb{Y}_{k,l}}\left[ i \right]}}} \right.} \right\}$ based on (\ref{probability2}).\\
                Compute $\Pr \left\{ {x\left[ {k,l} \right]|{\bf{Y}}} \right\}$ by using (\ref{sum_product}).\\
                Make hard decision of $x\!\left[ {k,\!l} \right]$ based on (\ref{MAP_rule}).
            }
        }
    }
}
\end{algorithm}
\vspace{-3mm}

\textbf{Remarks:}
Since the algorithm is derived from the \emph{symbol-wise} MAP sense, in principle, it is able to achieve the optimal error performance of OTFS systems in terms of the BER, if the corresponding graphical model does not contain any cycles.
Meanwhile, it can be observed from lines $5$ and $6$ of Algorithm~1 that the detection complexity of the \emph{symbol-wise} MAP algorithm is exponential to the number of paths $P$. However, such a complexity becomes prohibitive when the number of paths is significantly large. Therefore, we propose a reduced-complexity detection method based on the \emph{symbol-wise} MAP algorithm in the following section in order to strike a balance between the detection complexity and performance.

\section{Hybrid MAP and PIC Detection for OTFS Modulation}
It can be observed from~\eqref{sum_product} that the detection complexity mainly arises from the enumeration of all possible combinations of ${\mathbb{X}_{k,l}^{\left( i \right)}}$.
To reduce the detection complexity, we intend to separate the set ${\mathbb{X}_{k,l}^{\left( i \right)}}$ into two subsets and only enumerate the combinations of one subset.
Let $L$ be the size of the subset, whose total combinations are to be enumerated. Then, we have the following Proposition.

\textbf{Proposition 1} \emph{(Partitioning Rule)}:
Assuming that the path gains in $\mathbb{H}^{\left( i \right)}$ are sorted in descending order according to its power, i.e., $|h_k|^2>|h_j|^2$, if $k<j$, the two subsets of ${\mathbb{X}_{k,l}^{\left( i \right)}}$ are defined as
\begin{align}
\tilde{\mathbb{X}}_{k,l}^{\left( i \right)}&\buildrel \Delta \over = \left\{ {{\mathbb{X}}_{k,l}^{\left( i \right)}\left[ j \right]\left| {1 \le j \le L} \right.} \right\}
\end{align}
and
\begin{align}
\bar{\mathbb{X}}_{k,l}^{\left( i \right)}&\buildrel \Delta \over = \left\{ {{\mathbb{X}}_{k,l}^{\left( i \right)}\left[ j \right]\left| {L+1 \le j \le P-1} \right.} \right\},
\end{align}
respectively.
Naturally, we propose to perform MAP detection for the subset $\tilde{\mathbb{X}}_{k,l}^{\left( i \right)}$, while perform PIC for the subset $\bar{\mathbb{X}}_{k,l}^{\left( i \right)}$, since the subset $\bar{\mathbb{X}}_{k,l}^{\left( i \right)}$ may have a less impact on the overall error performance compared with that of the subset $\tilde{\mathbb{X}}_{k,l}^{\left( i \right)}$. More specifically, we assume that the elements in  $\bar{\mathbb{X}}_{k,l}^{\left( i \right)}$ are Gaussian variables \cite{Raviteja2018interference}, i.e., $\bar{\mathbb{X}}_{k,l}^{\left( i \right)}\left[ j \right]$ has a mean ${\mu _{k,l,i}}\left[ j \right]$ and variance $\sigma _{k,l,i}^2\left[ j \right]$.
In particular, the values of ${\mu _{k,l,i}}\left[ j \right]$ and $\sigma _{k,l,i}^2\left[ j \right]$ can be derived from the \emph{a posteriori} probabilities in the previous iteration.
Thus, we can modify (\ref{probability1}) as
\begin{align}
&\Pr \left\{ {{\mathbb{Y}_{k,l}}\left[ i \right]\left| {\tilde{\mathbb{X}}_{k,l}^{\left( i \right)},\bar{\mathbb{X}}_{k,l}^{\left( i \right)},x\left[ {k,l} \right]} \right.} \right\} = \frac{1}{{\sqrt {\pi \left( {{N_0} + {\sigma ^2}} \right)} }}\notag\\
&\exp \left( { - \left( {{\mathbb{Y}_{k,l}}\left[ i \right] - {h_i}x\left[ {k,l} \right] - \sum\nolimits_{j = 1}^L {{{\mathbb H}^{\left( i \right)}}\left[ j \right]\tilde{\mathbb{X}}_{k,l}^{\left( i \right)}[j]} } \right.} \right.\notag\\
&\left. {{{{{\left. { - \!\!\sum\nolimits_{j = 1}^{P - L - 1} \!\!{{{\mathbb H}^{\left( i \right)}}\!\left[ {j + L} \right]\mathbb{E}\left\{ \bar{\mathbb{X}}_{k,l}^{\left( i \right)}[j] \right\}} } \right)}^2\!}} \mathord{\left/
 {\vphantom {{{{\left. { - \!\!\sum\nolimits_{j = 1}^{P - L - 1}\!\! {{{\mathbb H}^{\left( i \right)}}\!\left[ {j + L} \right]\mathbb{E}\left\{ \bar{\mathbb{X}}_{k,l}^{\left( i \right)}[j] \right\}} } \right)}^2\!}} {\left( {{N_0} + {\sigma ^2}}\!\! \right)}}} \right.
 \!\kern-\nulldelimiterspace}\!\! {\left( {{N_0} + {\sigma ^2}} \right)}}} \right),
\label{probability3}
\end{align}
where ${\sigma ^2} = \sum\nolimits_{j = 1}^{P - L - 1} {\sigma _{k,l,i}^2\left[ j \right]} $.
Note that the messages passed between function nodes in the graphical model also need to be updated corresponding to (\ref{probability3}).
Specifically, for each function node ${\tilde{\mathbb{X}}_{k,l}^{\left( i \right)}}[j]$, the probability $\Pr \left\{ {{\tilde{\mathbb{X}}_{k,l}^{\left( i \right)}[j]}\left| {{\bf{Y}_{ \notin {\mathbb{Y}_{k,l}}\left[ i \right]}}} \right.} \right\}$ is passed to the function node ${\mathbb{Y}_{k,l}}\left[ i \right]$. On the other hand, for each function node ${\bar{\mathbb{X}}_{k,l}^{\left( i \right)}}[j]$, the corresponding mean ${\mu _{k,l,i}}\left[ j \right]$ and variance $\sigma _{k,l,i}^2\left[ j \right]$ are passed to the function node ${\mathbb{Y}_{k,l}}\left[ i \right]$;
The details of the proposed hybrid MAP and PIC detection algorithm are summarized in Algorithm~2.
\begin{algorithm}
\caption{Hybrid MAP and PIC Detection Algorithm for OTFS Modulation}
\KwIn{$\bf{Y}$, ${\mathbb{A}}$, $M$, $N$, $P$, the maximum number of iteration $I_{\max}$, \emph{a prior} probability $\Pr\left\{{x\left[ {k,l} \right]}\right\}$, and the channel state information $h_i$, ${l_\nu ^{\left( i \right)}}$, ${l_\tau ^{\left( i \right)}}$, for $1\le i \le P$.}
\KwOut{${\bf{\hat X}}$ and $\Pr \left\{ {\left. {x\left[ {k,l} \right]} \right|\bf{Y}} \right\}$.}
\For {$I=1;I \le I_{\max }$}
{
    \For {$i=1;i \le P$}
    {
        \For {$k=0;k \le N-1$}
        {
            \For {$l=0;l \le M-1$}
            {
                Calculate the mean and variance of each elements in $\bar{\mathbb{X}}_{k,l}^{\left( i \right)}$.\\
                Enumerate all combinations of $\tilde{\mathbb{X}}_{k,l}^{\left( i \right)}$.\\
                For each possible combination of $\tilde{\mathbb{X}}_{k,l}^{\left( i \right)}$, compute
                (\ref{probability3}) and $\Pr \left\{ {{\tilde{\mathbb{X}}_{k,l}^{\left( i \right)}[j]}\left| {{{\bf{Y}}_{ \notin {\mathbb{Y}_{k,l}}\left[ i \right]}}} \right.} \right\}$ based on (\ref{probability2}).\\
                Compute $\Pr \left\{ {x\left[ {k,l} \right]|{\bf{Y}}} \right\}$ by using (\ref{sum_product}).\\
                Make hard decision of $x\!\left[ {k,\!l} \right]$ based on (\ref{MAP_rule}).
            }
        }
    }
}
\end{algorithm}

\textbf{Remarks:} It can be observed that the detection complexity of the hybrid MAP and PIC detection algorithm is only exponential to $L$.
However, since PIC is applied in the algorithm, performance loss may be induced if the estimates of elements from $\bar{\mathbb{X}}_{k,l}^{\left( i \right)}$ are not accurate.
However, this performance loss is expected to be marginal for a coded OTFS system, since the channel code can usually provide reliable estimates of transmitted symbols. In general, iterations between the detector and channel-decoder are required in order to feed back useful information from the decoder to the detector.

On the other hand, we note that the message passing algorithm proposed in~\cite{Raviteja2018interference} is a special case of the proposed hybrid detection algorithm with $L=0$. Therefore, with $L>0$, the proposed hybrid detection algorithm can outperform the message passing algorithm in \cite{Raviteja2018interference}. In particular, with the increase of $L$, the error performance of the proposed hybrid detection algorithm can approach that of the \emph{symbol-wise} MAP algorithm introduced in Algorithm~1.

\vspace{-1mm}
\section{Numerical Results}
In this section, we investigate the error performance of the proposed algorithms under various channel conditions for both coded and uncoded OTFS modulation.
Without loss of generality, we set $N=100$ and $M=150$ for OTFS modulation, where the DD domain transmitted symbols are quadrature phase shift keying (QPSK) modulated. In order to demonstrate the advantage of the proposed algorithms, we also include the error performance of the message passing algorithm \cite{Raviteja2018interference} with a damping factor $0.7$ in our numerical results, where the number of iterations of the proposed algorithms and the message passing algorithm is set to be $I_{\rm{max}}=10$.

We set the maximum delay index as $l_{\max }=10$ and the maximum Doppler index as $k_{\max }=6$,
which is corresponding to a relative user equipment speed around $250$ km/h with $4$ GHz carrier frequency and $15$ kHz sub-carrier spacing \cite{Raviteja2018interference}. For each channel realization, we randomly select the
delay and Doppler indices such that $ - {k_{\max }} \le {l_\nu ^{\left( i \right)}} \le {k_{\max }}$ and $0 \le {l_\tau ^{\left( i \right)}} \le {l_{\max }}$.

\begin{figure}
\centering
\includegraphics[width=3in,height=2.5in]{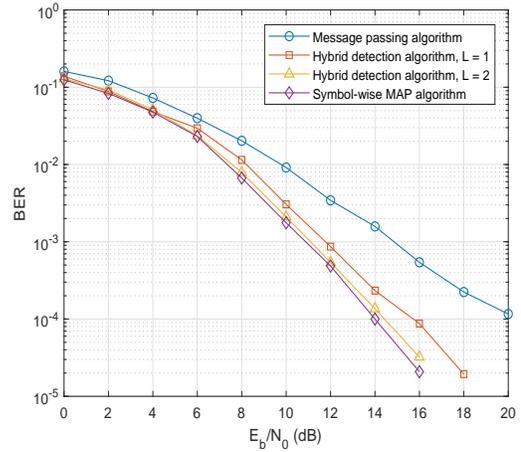}
\caption{BER performance of the proposed algorithms for uncoded OTFS systems with $P=4$, compared with the message passing algorithm in~\cite{Raviteja2018interference}.}
\label{uncoded}
\vspace{-4mm}
\centering
\end{figure}

Fig.~\ref{uncoded} demonstrates the BER performance of the proposed algorithms for uncoded OTFS systems with $P=4$. It can be observed from the figure that at BER $\approx 1 \times {10^{ - 4}}$, the proposed hybrid detection algorithm with $L=1$ shows a roughly $4.2$ dB gain compared with that of the message passing algorithm in~\cite{Raviteja2018interference}, which is consistent with our discussion.
Furthermore, we notice that the BER performance of the proposed hybrid detection algorithm improves with the increase of $L$.
More importantly, we observe that the BER performance of the proposed hybrid detection algorithm with $L=2$ approaches to that of the \emph{symbol-wise} MAP algorithm. This observation indicates that the proposed partitioning rule can effectively reduce the detection complexity without introducing a significant performance loss.

\begin{figure}
\centering
\includegraphics[width=3in,height=2.5in]{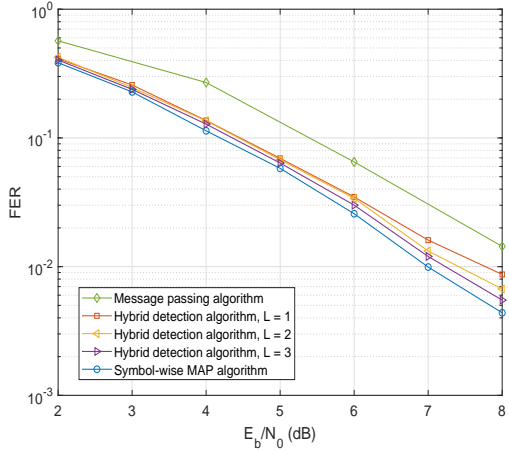}
\caption{FER performance of the proposed algorithms for Turbo coded OTFS systems with $P=5$.}
\label{Turbo_coded}
\vspace{-4mm}
\centering
\end{figure}

Fig.~\ref{Turbo_coded} demonstrates the frame error rate (FER) performance of the proposed algorithms for Turbo coded OTFS systems with $P=5$, where the
rate-$1/3$ Turbo code from the 3rd Generation Partnership Project (3GPP) wideband code division multiple access (WCDMA) standard~\cite{vucetic2012turbo} is applied.
We notice that the proposed hybrid detection algorithm shows a better FER performance with the increase of $L$ and noticeably outperforms the message passing algorithm in~\cite{Raviteja2018interference}. Meanwhile, it can be observed that with the application of channel coding, the proposed hybrid detection algorithm only has a marginal performance loss compared to the near-optimal \emph{symbol-wise} MAP algorithm even with $L=1$, where the performance loss is less than $1$ dB.
This observation indicates that our proposed algorithm can provide a good trade-off between the error performance and complexity.

\vspace{-1.5mm}

\section{Conclusion}
In this paper, we proposed a novel hybrid detection algorithm for OTFS modulation. We first derived the \emph{symbol-wise} MAP algorithm. Then, we proposed a partitioning rule which divides the related symbols into two sets according to their associated channel path gains.
A hybrid detection algorithm was proposed to exploiting the power discrepancy between the two subsets.
Specifically, the MAP detection was applied on the subset with larger channel gains, while the PIC detection was applied to the subset with smaller channel gains.
Simulation results verified the effectiveness our algorithm and showed that our proposed hybrid detection algorithm can provide a good trade-off between the error performance and detection complexity.

\vspace{-1.5mm}

\appendix
According to the Bayes's rule, (\ref{MAP_rule}) can be expanded as
\begin{equation}
\Pr \left\{ {x\left[ {k,l} \right]\left| {\bf{Y}} \right.} \right\} \propto \Pr \left\{ {{\bf{Y}}\left| {x\left[ {k,l} \right]} \right.} \right\}\Pr \left\{ {x\left[ {k,l} \right]} \right\} .
\label{derivation1}
\end{equation}
Let $\left. {{\mathbb{Y}_{k,l}}} \right|_{i + 1}^P$ denotes the vector of the $(i+1)$-th element ${\mathbb{Y}_{k,l}\left[ {i+1} \right]}$ to the $P$-th element ${\mathbb{Y}_{k,l}\left[ {P} \right]}$ of ${\mathbb{Y}_{k,l}}$.
By observing (\ref{DD_model}), (\ref{derivation1}) can be further derived according to the chain rule, which yields
\begin{align}
&\Pr \left\{ {{\bf{Y}}\left| {x\left[ {k,l} \right]} \right.} \right\}\Pr \left\{ {x\left[ {k,l} \right]} \right\}\notag\\
=& \prod\limits_{i = 1}^P {\Pr \left\{\! {{\mathbb{Y}_{k,l}}\left[ i \right]\left| {\left. {{\mathbb{Y}_{k,l}}} \right|_{i + 1}^P,\mathbb{Y}\backslash{\mathbb{Y}_{k,l}},x\left[ {k,l} \right]}\! \right.} \right\}} \Pr \left\{ {x\left[ {k,l} \right]} \right\},
\label{derivation2}
\end{align}
where ${\bf{Y}}\backslash{\mathbb{Y}_{k,l}}$ denotes the complementary set of ${\mathbb{Y}_{k,l}}$ with respect to $\bf{Y}$.
We further expand (\ref{derivation2}) as
\begin{align}
&\prod\limits_{i = 1}^P {\Pr \left\{ {{\mathbb{Y}_{k,l}}\left[ i \right]\left| {\left. {{\mathbb{Y}_{k,l}}} \right|_{i + 1}^P,{\bf{Y}}\backslash{\mathbb{Y}_{k,l}},x\left[ {k,l} \right]} \right.} \right\}} \Pr \left\{ {x\left[ {k,l} \right]} \right\}\notag\\
=&\prod\limits_{i = 1}^P {\sum\limits_{\mathbb{X}_{k,l}^{\left( i \right)}} {\!\Pr\! \left\{ \!{{\mathbb{Y}_{k,l}}\left[ i \right]\!,\mathbb{X}_{k,l}^{\left( i \right)}\!\left|\! {\left. {{\mathbb{Y}_{k,l}}} \right|_{i + 1}^P\!,\!{\bf{Y}}\backslash{\mathbb{Y}_{k,l}},x\left[ {k,l} \right]} \right.} \!\!\right\}} } \!\Pr\! \left\{ {x\left[ {k,l} \right]} \right\} \notag\\
=&\prod\limits_{i = 1}^P {\sum\limits_{\mathbb{X}_{k,l}^{\left( i \right)}} {\Pr \left\{ {{\mathbb{Y}_{k,l}}\left[ i \right]\left| {\left. {{\mathbb{Y}_{k,l}}} \right|_{i + 1}^P,{\bf{Y}}\backslash{\mathbb{Y}_{k,l}},{\mathbb{X}_{k,l}^{\left( i \right)}},x\left[ {k,l} \right]} \right.} \right\}} } \notag\\
&\quad\quad\quad\quad\Pr \left\{ {\left. {\mathbb{X}_{k,l}^{\left( i \right)}} \right|\left. {{\mathbb{Y}_{k,l}}} \right|_{i + 1}^P,{\bf{Y}}\backslash{\mathbb{Y}_{k,l}}} \right\}\Pr \left\{ {x\left[ {k,l} \right]} \right\}\label{derivation3}\\
=&\prod\limits_{i = 1}^P {\sum\limits_{{\mathbb{X}_{k,l}^{\left( i \right)}}} {\Pr \left\{ {{\mathbb{Y}_{k,l}}\left[ i \right]\left| {{\mathbb{X}_{k,l}^{\left( i \right)}},x\left[ {k,l} \right]} \right.} \right\}} } \notag\\
&\quad\quad\quad\quad\Pr \left\{ {\left. {\mathbb{X}_{k,l}^{\left( i \right)}} \right|\left. {{\mathbb{Y}_{k,l}}} \right|_{i + 1}^P,{\bf{Y}}\backslash{\mathbb{Y}_{k,l}}} \right\}\Pr \left\{ {x\left[ {k,l} \right]} \right\},
\label{derivation4}
\end{align}
where (\ref{derivation3}) is due to the Bayes's rule and the assumption that the information symbols in $\bf{X}$ are independent from each other.
Finally, by assuming that the elements from ${\mathbb{X}_{k,l}^{\left( i \right)}}$ are independent to the elements from $\left.{{\mathbb{Y}_{k,l}}} \right|_{1}^{i-1}$, we arrive at the conclusion given in Theorem 1. Note that the approximation becomes exact when the above assumption is valid, i.e., the corresponding graphical model does not contain any cycles.

\bibliographystyle{IEEEtran}
\bibliography{OTFS_references}

\end{document}